\documentclass[10pt,conference]{IEEEtran}
\usepackage{cite}
\usepackage{amsmath,amssymb,amsfonts}
\usepackage{algorithmic}
\usepackage{graphicx}
\usepackage{textcomp}
\usepackage[table,xcdraw]{xcolor}
\usepackage{booktabs}
\usepackage{multirow}
\usepackage{hyperref}
\usepackage{tcolorbox}
\def\BibTeX{{\rm B\kern-.05em{\sc i\kern-.025em b}\kern-.08em
    T\kern-.1667em\lower.7ex\hbox{E}\kern-.125emX}}

\newcommand{\uniba}{\textsc{Uniba}}
\newcommand{\upc}{\textsc{UPC}}

\begin{document}

\title{Teaching MLOps\\in Higher Education through Project-Based Learning
}

\author{

\IEEEauthorblockN{Filippo Lanubile}
\IEEEauthorblockA{
\textit{University of Bari}\\
Bari, Italy \\
filippo.lanubile@uniba.it}

\and

\IEEEauthorblockN{Silverio Martínez-Fernández}
\IEEEauthorblockA{
\textit{Universitat Politècnica de Catalunya}\\
Barcelona, Spain \\
silverio.martinez@upc.edu}

\and

\IEEEauthorblockN{Luigi Quaranta}
\IEEEauthorblockA{
\textit{University of Bari}\\
Bari, Italy \\
luigi.quaranta@uniba.it}

}


\maketitle


\begin{abstract}
Building and maintaining production-grade ML-enabled components is a complex endeavor that goes beyond the current approach of academic education, focused on the optimization of ML model performance in the lab.
In this paper, we present a project-based learning approach to teaching MLOps, focused on the demonstration and experience with emerging practices and tools to automatize the construction of ML-enabled components.
We examine the design of a course based on this approach, including laboratory sessions that cover the end-to-end ML component life cycle, from model building to production deployment.
Moreover, we report on preliminary results from the first edition of the course.
During the present year, an updated version of the same course is being delivered in two independent universities; the related learning outcomes will be evaluated to analyze the effectiveness of project-based learning for this specific subject.
\end{abstract}

\begin{IEEEkeywords}
machine learning, data science, software engineering for AI, model deployment, reproducibility\footnote{IEEE copyright notice (© 2023 IEEE): This is an Author’s Accepted Manuscript consisting of a post-peer-review, pre-copyedit version of a paper to be published by IEEE in the 2023 IEEE/ACM 45th International Conference on Software Engineering: Software Engineering Education and Training (ICSE-SEET). The final authenticated version will be available online.}
\end{IEEEkeywords}

\section{Introduction}
Over the past few years, Machine Learning (ML) techniques have been massively adopted in the software industry to solve a wide range of problems in a large and growing number of application domains \cite{deng2018artificial}.
As a direct consequence, ML-related job positions --e.g., data scientist, ML engineer, data engineer-- are among the most sought after in today's IT job market \cite{meesters2022ai}.

Notwithstanding its high promises, however, the adoption of ML typically turns out to be costly and challenging, especially for traditional software companies with little experience in the field. While building well-performing models is a relatively easy task to achieve -- especially thanks to modern ML frameworks, which abstract out the hard parts of the process -- integrating such models into production systems and maintaining them over time is a complex endeavor.
The results of an extensive survey published by Algorithmia in 2020 show that 55\% of the surveyed companies had not yet managed to deploy a single model to production, despite considerably investing in ML~\cite{algorithmia2020StateEnterprise2020}.

For this reason, several tools and practices are being proposed to support ML teams in translating their lab prototypes into production-grade ML components. For the most part, these activities are recognized as ``MLOps'', an umbrella term for the development of ML-enabled systems. Rooted in Software Engineering and inspired by DevOps \cite{lwakatareDevOps4AI}, MLOps places emphasis on process automation to achieve the continuous delivery of intelligent software functionalities; moreover, it takes into account the challenges that are peculiar to ML (e.g., reproducibility, explainability, and data privacy), and tries to address them with targeted improvements of the ML-component building process.

In the surge of enthusiasm for ML, many universities have started offering specialized degree courses and learning paths. Nonetheless, so far, the academic focus has been mainly on the construction of high-performance ML models, leveraging state-of-the-art techniques, such as deep learning~\cite{kastner2020teaching}. To the best of our knowledge, university courses do not generally emphasize MLOps, i.e., the practices and tools aimed at supporting the full life-cycle of production-ready ML components.
Besides delaying the acquisition of core competencies that are increasingly in demand, not providing training on MLOps represents a missed opportunity to fill the well-known cultural gap between data-related roles and software engineers.

In this paper, we describe our effort to teach MLOps practices and tools following a project-based learning method in the context of two university courses.
Specifically, these courses are offered by two independent universities located in different countries.
Both are targeted at students that are already acquainted with ML techniques; a key novelty is that the courses are based on a project assignment whose end goal is the deployment of a production-ready ML component.
Upon providing the students with basic training on a curated set of MLOps technologies, we ask them to leverage the presented tools to achieve their project end goal. Before and after project execution, we gauge the students' expectations and perceived instructional effectiveness of the adopted technologies.


\section{Related work}

With the proliferation of ML-based systems, notable research groups from large companies have highlighted the diverse skills and knowledge required for their development, deployment, and maintenance.
For instance, practitioners from Google were pioneers in bringing attention on the fact that software engineering practices shall not be dismissed for ML code \cite{sculley2015hidden}. In the other way around, practitioners from Microsoft Research remarked the growing demand of professionals with a deep knowledge of ML in ML-based systems projects \cite{simard2017machine}. In this context, it has emerged the need for a  multidisciplinary approach to develop ML-based systems, by involving data scientists (with a strong ML background), software engineers, and experts in the application domain. Indeed, this needed multidisciplinarity was argued in an SE4AI Ask-me-anything session as an educational problem~\cite{AMASoftwareEngineering2020}.
While there are many strong academic programs on either software engineering or data science/machine learning, their synergies are now increasingly starting to be key in academic curricula. In this scenario, we can find courses focused on the diverse skills required to build ML-based systems.

Kaestner and Kang created a course called ``Software engineering for AI-enabled systems'' for undergraduate to master and PhD students \cite{kastner2020teaching}. The course addresses fundamental SE skills for ML systems (requirements, design, quality assurance, operations, team, and processes) and responsible ML engineering (e.g., safety and ethics). This course is currently entitled ``Machine Learning in Production''. With high-quality theory contents and an online progressing book~\cite{kastnerMachineLearningProduction2021}, 
it does not apply project-based learning as a teaching methodology.

At a more abstract level, we can also find the need of offering new academic programs, aimed at training AI software engineers. Such a new role would cover both data science and software engineering. For instance, Bublin et al. propose a curriculum path for AI software engineers, which can be used to update a traditional SE curriculum
\cite{bublin2021educating}. In a similar direction, Heck et al. present a bachelor path for AI engineers \cite{heck2021lessons}. However, there is a gap of information and courses for teaching the whole MLOps lifecycle with a project-based learning method including clear guidelines of which practices and tools to apply. 

\section{The MLOps Project-Based Course}

We propose a project-based learning approach to teach MLOps.
The main objective of a course following this approach is to provide the students with theoretical and practical knowledge on building high-quality, production-grade ML components, ready to be integrated into ML-enabled systems.

The students will have an opportunity to train in both hard and soft skills.
First, they will form and consolidate expertise in MLOps; this will happen in the context of a practical project.
Second, they will strengthen collaboration and team-building skills, as they will work in groups of 3 to 5 people, leveraging agile practices such as tracking their work using a Kanban-style project board or conducting review and retrospective meetings at the end of each project milestone.
Third, the students will train their presentation skills, as teams will need to periodically present the progress of their work to the rest of the class.

The course is designed for both graduate and undergraduate students. A prerequisite is that students shall have a background in ML from previous courses (e.g., computer vision, speech recognition, etc.).
In particular, the attendees need to already know how to build an ML model from a data source.

\subsection{Project organization}

In the proposed course, the development of theoretical concepts goes hand in hand with their practical demonstration and application.
Throughout the course, the students engage with the execution of their project assignments.
Each theoretical lecture, introducing one or more MLOps practices, is followed by a practical demonstration of state-of-the-art MLOps tools.
Soon after these demos, the students have time to autonomously deepen the study of the topic and apply what they have learned to their own projects.

\subsection{Project milestones}

Overall, the project assignments are organized into 6 milestones with a 2-week duration, reflecting the distribution of the course contents.
In Table~I, for each milestone, we report the presented MLOps theory, the related practices, and the set of software solutions used to exemplify their implementation.
The criteria used by the instructors to select these tools are three-fold: their availability as open-source software, their popularity in the MLOps community, and their ease of learning.
It should be noted, however, that such choices are not binding for the students; indeed, the course attendees are free to explore available alternatives and pick their preferred option.

\begin{figure*}
  \centering
  \includegraphics[width=\textwidth]{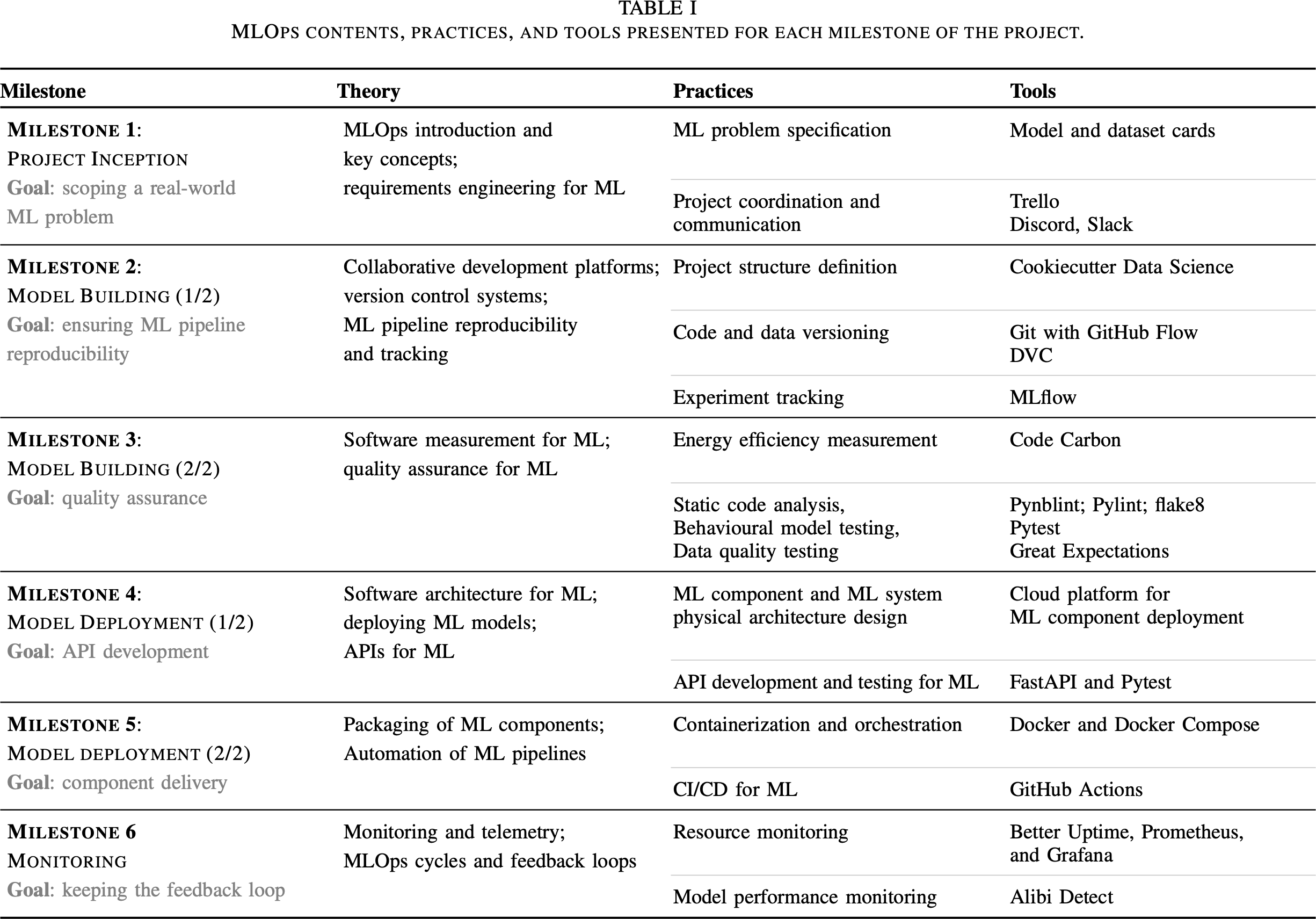}
\end{figure*}

\subsubsection{\textsc{Milestone 1} -- \textit{Project inception}}

the course starts with a general introduction to building ML-enabled systems, including key concepts and the main challenges.
Afterward, the instructors present the project assignment: the students will need to build a production-ready ML component that solves a real-world problem; the ML model on which the component will be based can be either pre-existing (e.g., created by the students to solve the assignment of a previous course) or newly developed in a parallel course.

\paragraph{ML problem specification}
as a first step, the members of each team discuss to select a problem that can be effectively solved with ML. Once the decision is made, they define the requirements for the ML component they intend to build;
these are documented by filling in a \textit{model card} and a \textit{dataset card}.
Model cards were introduced in \cite{mitchellModelCardsModel2019} to provide a structured description of ML models including their intended use cases, ethical considerations, and limitations; they provide a template for the reporting of basic model requirements and characteristics.
Together with \textit{dataset cards} -- an application of the same idea to the specification of datasets -- these templates are currently used to specify and organize ML assets within popular ML hubs like Hugging Face~\cite{HuggingFace}.

\paragraph{Project coordination and communication}
to finalize the project inception, the student teams set up a project management tool to coordinate their work and a communication channel to exchange information.

\subsubsection{\textsc{Milestone 2} -- \textit{Model building (ensuring ML pipeline reproducibility)}}

the focus of the second milestone is ensuring the reproducibility of the end-to-end model building process.

\paragraph{Project structure definition}
as a first step, the instructors underline the importance of consistently organizing ML projects in a structured way, to improve their readability and maintainability, facilitating collaboration among team members.
This can be achieved by sticking to a shared convention -- defined at the team level -- when organizing source code, data, and model files. To exemplify this practice, the instructors present the \textit{cookiecutter data science} template~\cite{drivendataCookiecutterDataScience}, a popular directory structure used to organize data science projects.

\paragraph{Code and data versioning}
then, the focus shifts to the versioning of source code and data.
Besides recommending the adoption of \texttt{git} as a Version Control System, the importance of sticking to a well-defined collaboration workflow, leveraging a code hosting platform like GitHub, is stressed. To this end, the instructors present the \textit{GitHub Flow} workflow~\cite{githubGitHubFlow}, a popular branch-based convention for organizing the development of software projects on GitHub. Also, the intricacies of versioning and comparing computational notebooks (e.g., Jupyter notebooks) are discussed, including approaches to overcome the current limitations of this process.

ML models result from the application of algorithms to data; therefore, data is a first citizen in ML workflows and -- for the sake of model traceability and reproducibility -- its versioning is as important as the versioning of source code.
To show how the data versioning process can be seamlessly integrated with code versioning, the instructors demonstrate the use of DVC (``Data Version Control''), a popular open-source solution that integrates with \texttt{git} to track the versions of large data files. DVC can also be used to memorize the computational steps that compose an ML pipeline, keeping track of the dependencies and outputs of each step and -- ultimately -- codifying a reproducible ML workflow in a human-readable YAML file.

\paragraph{Experiment tracking}
to ensure the reproducibility and accountability of ML model building pipelines and record the rationale behind the several decisions involved, another important aspect to consider is the tracking of ML experiments.
Indeed, during fast paced experimentations, it is easy to lose track of the various trials performed, i.e., of how variations of the datasets and changes in the experimental configuration
affect the performance of output models.
To exemplify experiment tracking, we use the Tracking module of MLflow \cite{noauthor_mlflow_nodate}, a popular open-source platform for the management of the end-to-end ML lifecycle.
MLflow features an intuitive Python API that allows its users to log the parameters and metrics of each experiment, as well as the artifacts produced by the ML pipeline (e.g., the pre-processed dataset used for model training and the trained ML models); after each experiment execution, the logged information can be visualized and compared using a web UI.

\subsubsection{\textsc{Milestone 3} -- \textit{Model building (quality assurance)}}

the third project milestone is focused on assuring the quality of the model building pipeline and the resulting models.

\paragraph{Energy efficiency measurement}

besides requiring substantial human effort, training state-of-the-art ML models is a resource-demanding process; computationally intensive algorithms are repeatedly executed on large input datasets to train ML models and keep them updated over time.
For this reason, the energy efficiency of model training pipelines should be taken into due account and considered as an important quality factor of ML-enabled components \cite{SchwartzGreenAI, StrubellEnergyAndPolicy}.
To sensitize the students to this theme and provide them with a practical solution to assess their training pipelines, we demonstrate the use of Code Carbon; shipped as a lightweight Python package, this tool can be employed to estimate the amount of carbon dioxide ($CO_{2}$) produced by various kinds of computing resources.

\paragraph{Static analysis and testing}

the source code developed during the experimental phase of ML projects constitutes the foundation of production code. Hence, to facilitate the translation of experimental artifacts (i.e., of scripts and computational notebooks) into production-grade components, it is crucial to 
adopt high quality standards since the start of ML projects.
Towards this aim, practitioners generally implement well-established software engineering practices like static program analysis and testing; however, these are not always easy to adapt and apply to experimental ML code: for instance, assuring the quality of computational notebooks is not straightforward. For this reason, along with popular static analysis tools of the Python ecosystem (e.g., Pylint and flake8), we present and demonstrate the use of Pynblint \cite{quarantaPynblintStaticAnalyzer2022}, a specialized linter for notebooks in the \texttt{.ipynb} format.
Based on a catalog of empirically validated best practices \cite{quarantaElicitingBestPractices2022}, Pynblint can be employed to analyze Jupyter notebooks written in Python and detect potential defects hindering the reproducibility of computations as well as collaboration among team members.

Another crucial aspect of QA is testing.
As the operation of ML-enabled components is not only based on code, but also on data and models, all of these entities need to be thoroughly tested to ensure the quality of ML-enabled systems.
Accordingly, we demonstrate to the students how Pytest \cite{noauthor_pytest_nodate} can be adopted to test code and models while Great Expectations \cite{noauthor_great_nodate} is employed for data testing.
Pytest is a general-purpose testing framework for Python. It can be used to test model training pipelines as well as the performance and behavior of models.
As for the latter, we emphasize the importance to gauge the quality of models not only based on quantitative metrics (e.g., accuracy, precision, and recall), but also in terms of the behavior exhibited when models are applied to specific categories of input data.
Behavioral model testing is a novel research field addressing this need; 
in our course, we introduce the students to this practice by means of examples -- implemented with Pytest -- of behavioral test instances drawn from ~\cite{ribeiroAccuracyBehavioralTesting2020}.
Finally, to test the quality of data, we demonstrate the use of Great Expectations; this specialized testing framework 
offers dozens of pre-defined tests and various utilities, like data profilers. Moreover, it can be used to document the data sources used in an ML project.

\subsubsection{\textsc{Milestone 4} -- \textit{Model deployment (API development)}}

once an ML model has been trained and fine-tuned, it shall be deployed as an ML component to be easily integrated in a larger ML system.

\paragraph{ML component and system physical architecture design}

we start by examining the most common architectural styles used to organize the components of ML-enabled systems: N-tier, microservices, batch, and stream big data architectures. Students shall understand the impact that deploying an ML component to the server or client side has on quality attributes.
Upon designing the physical architecture of their component, the students select a cloud platform to host it.
Beyond the variegated set of commercial options -- which typically offer free tiers that are sufficient to host proof-of-concept applications -- the students can leverage virtual machines provided by their own university.

\paragraph{API development and testing for ML}

then, we delve into the development of APIs for ML services.
Specifically, we present background information on the design of web-based software interfaces and characterize the most common architectural styles (e.g., RPC and REST).
Next, to provide the students with practical guidance on this matter, we demonstrate the implementation of a RESTful API for a simple ML model.
The featured example is developed using FastAPI \cite{noauthor_fastapi_nodate}, an open-source web framework for Python; FastAPI allows the development and documentation of production-grade APIs that comply with OpenAPI (i.e., the industry-standard specification); moreover, the framework offers testing utilities that can be used with Pytest to verify API endpoints.

\subsubsection{\textsc{Milestone 5} -- \textit{Model deployment (component delivery)}}

the next milestone is dedicated to the delivery of ML components as self-contained software packages.

\paragraph{Containerization and orchestration}

to date, containerization is one of the most widespread forms of software packaging.
Using containers to ship ML components has several advantages, e.g., the portability and reproducibility of the execution environment, which is fully specified in code.
Furthermore, thanks to container orchestrators -- i.e., programs designed to automate the end-to-end container life cycle -- container-based systems tend to be robust to failures and easy to scale.
In this regard, we demonstrate the use of Docker \cite{noauthor_docker_2022}, the de-facto standard for software containerization, and of Docker Compose,
a command-line tool for the orchestration of multi-container Docker applications.

\paragraph{CI/CD for ML}

process automation is one of the main prerogatives of MLOps.
Thus, to refine the deployment process, we explain how the execution of the main MLOps practices presented so far can be fully automatized in the context of Continuous Integration and Continuous Delivery (CI/CD) workflows (from model training, to quality assurance and containerization).
In particular, to exemplify this practice we leverage GitHub Actions, a workflow automation tool integrated with GitHub.

\subsubsection{\textsc{Milestone 6} -- \textit{Monitoring}}

the final project milestone is dedicated to the monitoring of deployed ML components.

\paragraph{Resource monitoring} first, we stress the importance of continually monitoring the performance and resource consumption of ML-components seen as standard software units; the metrics collected as part of this practice enable operations teams to make timely decisions concerning resource allocation; moreover, the same information can be leveraged to automatically scale resources and balance the load within cloud-based execution environments.
The minimum form of resource monitoring is availability monitoring. In our course, we exemplify this practice using Better Uptime \cite{noauthor_better_nodate}, a commercial availability monitor operated via a user-friendly web application.
Then, we demonstrate how a couple of open-source solutions can be used, in tandem, to collect and visualize comprehensive sets of resource key performance indicators (KPIs): Prometheus \cite{prometheus_prometheus_nodate} -- an open-source monitoring system based on an efficient time series database -- and Grafana \cite{noauthor_grafana_nodate} -- an open-source dashboarding software for live data visualization.

\paragraph{Model performance monitoring} 

no matter how accurate models are, the aspects of reality that they capture are expected to change over time; this is at the root of well-known phenomena like data and concept drift, which in turn entail the degradation of performances in production models.
The early detection of these issues is a crucial aspect of the maintenance of ML-enabled components. For this reason, the input data stream of each model should be continuously monitored to timely detect drifts and trigger model re-training pipelines.
In our course, we demonstrate how the Alibi Detect library~\cite{AlibiDetect} can be used to periodically monitor production data and determine the emergence of outliers, alterations in variable distributions, and a range of further anomalies that might result in worsened model performances.

\section{Proposed formal evaluation strategy}

In 2022 Fall, the project-based learning course is going to be offered by two independent institutions, i.e., the University of Bari (Italy) -- hereinafter \uniba{} -- and the Universitat Politècnica de Catalunya (Spain) -- hereinafter \upc.
In particular, \uniba{} will offer the course to graduate students attending an MSc in Computer Science, while \upc{} will deliver it to undergraduate students from a BSc in Data Science.
In both cases, the students will be already well trained on building ML models, having succeeded in previous courses on the subject. We use the Goal/Question/Metric (GQM) method \cite{Basili94} as an aid to shape the evaluation goal.

\begin{tcolorbox}[colback=gray!5!white,colframe=gray!75!black]
\textit{Analyze} the project-based learning approach to teach MLOps 
\textit{for the purpose of} evaluation
\textit{with respect to} the instructional effectiveness 
\textit{from the point of view of} instructors and students 
\textit{in the context of} university courses.
\end{tcolorbox}

In particular, we will measure instructional effectiveness by focusing on two learning outcomes~\cite{GuoAReviewOf}:
\subsubsection{\textit{Affective outcomes}} i.e., how the students perceive the benefits and, more in general, the experience of project-based learning. 

We will collect qualitative feedback from the students by administering one or more surveys.
For instance, the questionnaires will gauge whether the students perceived working in a team at the lab project useful to learn MLOps.

\subsubsection{\textit{Artifact quality}} i.e., the quality of the artifacts delivered by the students at each milestone.

We will assess the quality of the artifacts produced by the students by looking at how they applied each of the MLOps practices from Table~I. Specifically, we will consider three possible outcomes:
\begin{itemize}
\item {\textit{poor}}: the students were not able to apply the recommended MLOps practice;
\item {\textit{fair}}: the students implemented the MLOps practice by replicating the provided examples with minor changes;
\item {\textit{good}}: the students implemented the MLOps practice in an extended or innovative way.
\end{itemize}

\section{Preliminary results}

In 2021 Fall, we offered a pilot version of the proposed course to graduate students at \uniba{}. Although there was not a formal evaluation strategy for the pilot, we had a chance to appreciate the positive feedback from the students and learn some lessons from the course retrospective. First, we learned that we could avoid some tutorials because the contents were already known (e.g. Jupyter notebooks and \texttt{git}) or felt as overkill for the task (e.g., Kubernetes).
On the other side, we realized the need to go beyond resource monitoring to also include model performance monitoring and then teach how to notice the model drift in production.

By the end of the course, all the student teams were able to deploy their ML components in a cloud-based production environment. Most of them found the proposed technologies attractive and useful to carry out the project assignments; in some cases, they even went beyond the instructor demonstrations, trying out alternative software solutions (e.g., TensorBoard instead of MLflow for experiment tracking). Finally, the presentations conducted at the end of each milestone were perceived by the students as inspiring opportunities to learn alternative solutions to the challenges faced.

\section{Conclusion}

The topics of the project-based course on MLOps cover the end-to-end life cycle of ML-enabled software components; for each MLOps practice presented in theory, an hands-on demonstration of its practical application is delivered to the students, who in turn apply the acquired knowledge to build and deploy their own ML-enabled component.
As future work, we will analyze the effectiveness of the proposed project-based learning approach by taking into account the student feedback and the quality of their work.

\section*{Acknowledgment}
This work is partially supported by the project TED2021-130923B-I00, funded by MCIN/AEI/10.13039/501100011033 and the European Union Next Generation EU/PRTR.

\bibliographystyle{IEEEtran}
\bibliography{IEEEabrv,ICSE-SEET-2023-refs}

\end{document}